\documentclass[journal]{IEEEtran}
\usepackage[latin1]{inputenc}
\usepackage{times}
\usepackage{graphicx}
\usepackage{amsmath}
\usepackage{amssymb}
\usepackage{nicefrac}
\usepackage{psfrag}
\usepackage{multirow}
\usepackage{mathptmx}

\newcommand{\ISL}{{\rm ISL}}

\renewcommand{\Re}[1]{{\rm Re}\left\{#1\right\}}
\newcommand{\iu}{{\mathbf{i}}}
\newcommand{\XX}{{\mathbb{X}}}

\newcommand{\aseq}{\sim}
\newcommand{\umod}[1]{\left[#1\right]_1}

\begin{document}

\title{Determination of the Integrated Sidelobe Level of Sets of Rotated Legendre Sequences}

\author{Javier~Haboba,~\IEEEmembership{Student Member,~IEEE,}
        Riccardo~Rovatti,~\IEEEmembership{Member,~IEEE,}
        and~Gianluca~Setti,~\IEEEmembership{Member,~IEEE}}
        

\maketitle

\begin{abstract}
Sequences sets with low aperiodic auto- and cross-correlations play an important role in many applications like communications, radar and other active sensing applications. The use of antipodal sequences reduces hardware requirements while increases the difficult of the task of signal design. In this paper we present a method for the computation of the Integrated Sidelobe Level ($\ISL$), and we use it to calculate the asymptotic expression for the $\ISL$ of a set of sequences formed by different rotations of a Legendre sequence.
\end{abstract}


\section{Introduction}
\label{sec:intro}

\IEEEPARstart{T}{he} design of sequences set with good correlation properties is present in many fields of engineering such as radar, sonar, communications, medical imaging and so on. Good auto-correlation properties means that any sequence in the set is nearly uncorrelated with its own shifted version while good cross-correlation means that any member of the sequences set is nearly uncorrelated with any other members at any shift.

A commonly used metric of the goodness of the correlation is the \textit{Integrated Sidelobe Level} ($\ISL$).
The $\ISL$ of a set of $M$ sequences each of $N$ (possibly complex) symbols that we will indicate with $x^{(p)}_j$ with $j = 0,\dots, N-1$ and $p = 0,\dots, M-1$ is defined as

\begin{equation*}
\ISL =  \sum_{p=0}^{M-1}\sum_{\substack{k=-N+1\\k\neq 0}}^{N-1}
\left|X_{x^{(p)}x^{(p)}}(k)\right|^2 
 +\sum_{p=0}^{M-1}\sum_{\substack{q=0\\p\neq q}}^{M-1} \sum_{k=-N+1}^{N-1}
|X_{x^{(p)}x^{(q)}}(k)|^2
\end{equation*}

\noindent where

\begin{align*}
X_{x^{(p)}x^{(p)}}(k)=\sum_{j=\max\{0,-k\}}^{\min\{N-k,N\}-1} x^{(p)}_j x^{*(p)}_{j+k}
&&
k = -N+1 \dots N-1
\end{align*}

\noindent is is the auto-correlation of the sequence $\textbf{\textit{x}}^{(p)}$, and

\begin{align*}
X_{x^{(p)}x^{(q)}}(k)=\sum_{j=\max\{0,-k\}}^{\min\{N-k,N\}-1} x^{(p)}_j x^{*(q)}_{j+k}
&&
k = -N+1 \dots N-1
\end{align*}

\noindent is the cross-correlation between the sequences $\textbf{\textit{x}}^{(p)}$ and $\textbf{\textit{x}}^{(q)}$.

Good set of sequences are those having a low $\ISL$ value.
Due to the strong interest in the design of sequences with low $\ISL$ value, many algorithms have been suggested for its minimization. Our purpose is to develop an analytical expression that may drive optimization in some particular difficult cases, most notably when the antipodal constrain $(x^{p}_j=\pm 1)$ is imposed.

To facilitate the discussion, denote the sum of squares corresponding to the auto-correlation terms as

\begin{equation}
\label{eq:XXaa}
\XX_{x^{(p)}x^{(p)}} = \sum_{\substack{k=-N+1\\k\neq 0}}^{N-1}\left|X_{x^{(p)}x^{(p)}}(k)\right|^2
\end{equation}

\noindent and the sum of squares corresponding to the cross-correlation terms as

\begin{align}
\label{eq:XXab}
\XX_{x^{(p)}x^{(q)}} = \sum_{k=-N+1}^{N-1} |X_{x^{(p)}x^{(q)}}(k)|^2 && p \neq q
\end{align}

\noindent so that

\begin{equation}
\label{eq:ISL_sim}
\ISL =  \sum_{p=0}^{M-1}\XX_{x^{(p)}x^{(p)}}
 +\sum_{p=0}^{M-1}\sum_{\substack{q=0\\p\neq q}}^{M-1} \XX_{x^{(p)}x^{(q)}}
\end{equation}

A general method for the calculation of $\XX_{x^{(p)}x^{(p)}}$ of any sequences of odd length is presented in \cite{Hoholdt_1988,Hoholdt_2006}. This method hinges on generating functions and writes correlations as proper sums of their values on the unit circle in the complex plane.
The method works well when we have analytical insights on the generating functions.


Extending the ideas of \cite{Hoholdt_1988}, in section \ref{sec:calc} we devise a general method for the calculation of $\XX_{x^{(p)}x^{(q)}}$ in \eqref{eq:XXab} of any pair of real sequences of odd length and thus, together with the result in \cite{Hoholdt_1988,Hoholdt_2006}, the $\ISL$ for a set of sequences. In section \ref{sec:LegSec} we use this method to obtain an asymptotic expression for the $\ISL$ value of a set formed by different rotations of Legendre sequences.
Some minor results about an optimization procedure based on the latter expression are reported in \cite{Haboba_2010}, where we  find the optimal rotations that minimize the $\ISL$ for any sequences length N.

Throughout the paper we use the following asymptotic notation.

We say that

\begin{itemize}

\item

two sequences $a_N$ and $b_N$ are asymptotically equivalent, $a_N \sim b_N$ iff

\begin{equation*}
	\mathop {\lim }\limits_{N \to \infty} \frac{a_N}{b_N} = 1
\end {equation*}

\item

$a_N$ is asymptotically bounded by $b_N$, $a_N = O(b_N)$ iff

\begin{align*}
	\exists M>0 \text{ and } \exists N_o && \big| && \left| a_N \right| \leq M \left| b_N \right| && \forall N > N_o
\end {align*}

\end{itemize}

\section{Calculation of the cross-correlation terms in the $\ISL$}
\label{sec:calc}

Let $a_0,a_1, \dots, a_{N-1}$ and $b_0,b_1, \dots, b_{N-1}$ be two real sequences of length N, we want to obtain an expression for $\XX_{ab}$.

If we define the generating functions of the two sequences as

\begin{align*}
Q_a(z)&=\sum_{j=0}^{N-1}a_j z^j &
Q_b(z)&=\sum_{j=0}^{N-1}b_j z^j
\end{align*}

\noindent we have that

\begin{equation*}
Q_a(z)Q^*_b(z)=\sum_{k=-N+1}^{N-1}X_{ab}(k)z^{-k}
\end{equation*}

\noindent and thus

\begin{equation*}
\left|Q_a(z)Q^*_b(z)\right|^2=\sum_{k=-N+1}^{N-1}\sum_{l=-N+1}^{N-1}X_{ab}(k)X_{ab}(l)z^{-k+l}
\end{equation*}

Now, set $\epsilon_j=e^{\frac{2\pi\iu}{N}j}$ and note that for $k,l=-N+1,\dots,N-1$,

\begin{equation*}
\sum_{j=0}^{N-1}\epsilon_j^{-k+l}=\begin{cases}
N & \text{if $-l+k=-N,0,N$}\\
0 & \text{otherwise}
\end{cases}
\end{equation*}

Hence, if we define 

\begin{multline*}
S'
=
\sum_{j=0}^{N-1}\left|Q_a(\epsilon_j)Q^*_b(\epsilon_j)\right|^2
 = N\sum_{k=-N+1}^{N-1}X^2_{ab}(k)+\\
   N\sum_{k=1}^{2N-1}X_{ab}(k)X_{ab}(k-N)+
   N\sum_{k=-N+1}^{-1}X_{ab}(k)X_{ab}(k+N)\notag
\end{multline*}
\noindent and (for $N$ odd)
\begin{multline*}
S''
=
\sum_{j=0}^{N-1}\left|Q_a(-\epsilon_j)Q^*_b(-\epsilon_j)\right|^2
 = N\sum_{k=-N+1}^{N-1}X^2_{ab}(k)+\\
  -N\sum_{k=1}^{2N-1}X_{ab}(k)X_{ab}(k-N)
  -N\sum_{k=-N+1}^{-1}X_{ab}(k)X_{ab}(k+N)\notag
\end{multline*}

\noindent we can express $\XX_{ab}$ (i.e. the sum of squares of cross-correlations as in \eqref{eq:XXab}) as

\begin{equation*}
\XX_{ab}=\sum_{k=-N+1}^{N-1}X^2_{ab}(k)=
\frac{S'+S''}{2N}
\end {equation*}

To compute $S''$ we use the Lagrange interpolation polynomials to calculate the values of $Q_a(-\epsilon_j)$ from $Q_a(\epsilon_k)$ for $j,k=0, \dots, N-1$. In this special case the data points ($\epsilon_k$) coincide with the complex roots of unity and, for $N$ odd, the Lagrange base polynomials simply reduce to $\frac{2}{N}\frac{\epsilon_k}{\epsilon_j+\epsilon_k}$ \cite[p. 89]{Polya_1925}. Then

\begin{equation}
\label{eq:interp}
Q_a(-\epsilon_j)=\frac{2}{N}\sum_{k=0}^{N-1}\frac{\epsilon_k}{\epsilon_j+\epsilon_k}Q_a(\epsilon_k)
\end{equation}

By substituting \eqref{eq:interp} into $S''$ and developing the product $\left|Q_a(-\epsilon_j)Q^*_b(-\epsilon_j)\right|^2$ we get

\begin{align*}
S''
&=&
\frac{16}{N^4}
\sum_{j=0}^{N-1}
\left[
\sum_{k_1=0}^{N-1}
\frac{\epsilon_{k_1}}{\epsilon_j+\epsilon_{k_1}}
Q_a(\epsilon_{k_1})
\sum_{l_1=0}^{N-1}
\frac{\epsilon^*_{l_1}}{\epsilon^*_j+\epsilon^*_{l_1}}
Q^*_a(\epsilon_{l_1})\right.\\
&&
\left.
\sum_{k_2=0}^{N-1}
\frac{\epsilon_{k_2}}{\epsilon_j+\epsilon_{k_2}}
Q_b(\epsilon_{k_2})
\sum_{l_2=0}^{N-1}
\frac{\epsilon^*_{l_2}}{\epsilon^*_j+\epsilon^*_{l_2}}
Q^*_b(\epsilon_{l_2})
\right]\\
&=&
\frac{16}{N^4}
\sum_{k_1=0}^{N-1}
\sum_{l_1=0}^{N-1}
\sum_{k_2=0}^{N-1}
\sum_{l_2=0}^{N-1}
Q_a(\epsilon_{k_1})
Q^*_a(\epsilon_{l_1})
Q_b(\epsilon_{k_2})
Q^*_b(\epsilon_{l_2})\\
&&
\sum_{j=0}^{N-1}
\frac{\epsilon_{k_1}}{\epsilon_j+\epsilon_{k_1}}
\frac{\epsilon^*_{l_1}}{\epsilon^*_j+\epsilon^*_{l_1}}
\frac{\epsilon_{k_2}}{\epsilon_j+\epsilon_{k_2}}
\frac{\epsilon^*_{l_2}}{\epsilon^*_j+\epsilon^*_{l_2}}
\end{align*}

\noindent in which we may exploit the fact that $\epsilon^*_j=1/\epsilon_j$ to write

\begin{align}
\label{eq:S2}
S''=
\frac{16}{N^4}
\sum_{k_1=0}^{N-1}
\sum_{l_1=0}^{N-1}&
\sum_{k_2=0}^{N-1}
\sum_{l_2=0}^{N-1}
\epsilon_{k_1}
\epsilon_{k_2}
Q_a(\epsilon_{k_1})
Q^*_a(\epsilon_{l_1})
Q_b(\epsilon_{k_2})
Q^*_b(\epsilon_{l_2})\notag\\
&
\times
\bigg\{
\sum_{j=0}^{N-1}
\frac{1}{\epsilon_j+\epsilon_{k_1}}
\frac{\epsilon_j}{\epsilon_j+\epsilon_{l_1}}
\frac{1}{\epsilon_j+\epsilon_{k_2}}
\frac{\epsilon_j}{\epsilon_j+\epsilon_{l_2}}
\bigg\}
\end{align}

Let us define now the innermost sum of \eqref{eq:S2} as

\begin{align*}
W(k_1,l_1,k_2,l_2)
&=
\sum_{j=0}^{N-1}
\frac{1}{\epsilon_j+\epsilon_{k_1}}
\frac{\epsilon_j}{\epsilon_j+\epsilon_{l_1}}
\frac{1}{\epsilon_j+\epsilon_{k_2}}
\frac{\epsilon_j}{\epsilon_j+\epsilon_{l_2}}\\
&=
\sum_{j=0}^{N-1}f_{k_1,l_1,k_2,l_2}(\epsilon_j)
\end{align*}

\noindent with
\begin{equation*}
f_{p,q,r,s}(z)=\frac{z^2}{(z+\epsilon_{p})(z+\epsilon_{q})(z+\epsilon_{r})(z+\epsilon_{s})}
\end{equation*}

Depending on $p,q,r,s$, the rational function $f_{p,q,r,s}(z)$ can be
transformed into a specific sum of simple rational parts. Each of these
rational parts can be summed separately.  This path is fully
developed in \cite{Hoholdt_1988} and we here exploit the results
therein.

In particular we have that

\begin{itemize}

\item[$A)$] for $0\le p<N$

\begin{equation*}
W(p,p,p,p)=\frac{1}{16}\left(\frac{1}{3}N^4+\frac{2}{3}N^2\right)\frac{1}{\epsilon_p^2}
\end{equation*}

\item[$B)$] for $0\le p\neq q<N$

\begin{align*}
&W(p,p,p,q)=W(p,p,q,p)=W(p,q,p,p)=\\
&W(q,p,p,p)
=
\frac{1}{8}N^2
\left(\frac{\epsilon_q+\epsilon_p}{\epsilon_p(\epsilon_q-\epsilon_p)^2}
\right)
\end{align*}

\item[$C)$] for $0\le p\neq q\neq r<N$

\begin{align*}
\lefteqn{W(p,p,q,r) = W(p,p,r,q)=W(p,q,p,r)=}\\
&W(p,r,p,q) = W(p,q,r,p)=W(p,r,q,p)=\\
&W(q,p,r,p) = W(r,p,q,p)=W(q,r,p,p)=\\
&W(r,q,p,p) = -\frac{1}{4}N^2\frac{1}{\epsilon_q-\epsilon_p}\frac{1}{\epsilon_r-\epsilon_p}
\end{align*}

\item[$D)$] for $0\le p\neq q<N$

\begin{align*}
&W(p,p,q,q)=W(p,q,p,q)=W(p,q,q,p)=\\
& -\frac{1}{2}N^2\frac{1}{(\epsilon_p-\epsilon_q)^2}
\end{align*}

\item[$E)$] for $0\le p\neq q\neq r\neq s<N$

\begin{equation*}
W(p,q,r,s)=0
\end{equation*}

\end{itemize}

Taking into account all the above cases we may write $S''=\frac{16}{N^4}\left(\alpha+\beta+\gamma+\delta\right)$, where the terms $\alpha$, $\beta$, $\gamma$, and $\delta$ correspond to the contributions of the cases A, B, C and D respectively.

For the cases included in $A)$ we have that

\begin{align}
\label{eq:A}
\alpha
=
\frac{1}{16}\left(\frac{1}{3}N^4+\frac{2}{3}N^2\right)
\sum_{p=0}^{N-1}\left|Q_a(\epsilon_p)Q_b(\epsilon_p)\right|^2
\hspace{17mm}
\end{align}

\noindent for the cases in $B)$ we have

\begin{align}
\label{eq:B}
\beta
=
\frac{1}{8}N^2\sum_{\substack{p,q=0\\p\neq q}}^{N-1}
\Biggl\{&
\left(
\frac{\epsilon_q+\epsilon_p}{\epsilon_p(\epsilon_q-\epsilon_p)^2}
\right)
\times
\\
&
\Bigl[\epsilon_p^2
\left|Q_a(\epsilon_p)\right|^2Q_b(\epsilon_p)Q^*_b(\epsilon_q)+
\nonumber \\
&\hspace{10mm}
\epsilon_{p}\epsilon_{q}\left|Q_a(\epsilon_p)\right|^2Q_b(\epsilon_q)Q^*_b(\epsilon_p)+
\nonumber \\
&\hspace{15mm}
\epsilon_{p}^2Q_a(\epsilon_p)Q^*_a(\epsilon_q)\left|Q_b(\epsilon_p)\right|^2+
\nonumber \\
&\hspace{20mm}
\epsilon_{q}\epsilon_{p}Q_a(\epsilon_q)Q^*_a(\epsilon_p)\left|Q_b(\epsilon_p)\right|^2
\Bigr]\Biggr\}
\nonumber
\end{align}

\noindent for $C)$ we have

\begin{align}
\label{eq:C}
\gamma
=
\frac{1}{4}N^2
\sum_{\substack{p,q,r=0\\p\neq q\neq r}}^{N-1}
\Biggl\{&
\frac{-1}{(\epsilon_q-\epsilon_p)(\epsilon_r-\epsilon_p)}\times\\
&
\Bigl[
\epsilon_{p}
\epsilon_{q}
\left|Q_a(\epsilon_p)\right|^2
Q_b(\epsilon_{q})
Q^*_b(\epsilon_{r})
+
\nonumber \\
& \hspace{2mm}
\epsilon_{p}
\epsilon_{r}
\left|Q_a(\epsilon_{p})\right|^2
Q_b(\epsilon_{r})
Q^*_b(\epsilon_{q})
+
\nonumber \\
& \hspace{4mm}
\epsilon_{p}^2
Q_a(\epsilon_{p})
Q^*_a(\epsilon_{q})
Q_b(\epsilon_{p})
Q^*_b(\epsilon_{r})
+
\nonumber \\
& \hspace{6mm}
\epsilon_{p}^2
Q_a(\epsilon_{p})
Q^*_a(\epsilon_{r})
Q_b(\epsilon_{p})
Q^*_b(\epsilon_{q})
+
\nonumber \\
& \hspace{8mm}
\epsilon_{p}
\epsilon_{r}
Q_a(\epsilon_{p})
Q^*_a(\epsilon_{q})
Q_b(\epsilon_{r})
Q^*_b(\epsilon_{p})
+
\nonumber \\
& \hspace{10mm}
\epsilon_{p}
\epsilon_{q}
Q_a(\epsilon_{p})
Q^*_a(\epsilon_{r})
Q_b(\epsilon_{q})
Q^*_b(\epsilon_{p})
+
\nonumber \\
& \hspace{12mm}
\epsilon_{q}
\epsilon_{r}
Q_a(\epsilon_{q})
Q^*_a(\epsilon_{p})
Q_b(\epsilon_{r})
Q^*_b(\epsilon_{p})
+
\nonumber \\
& \hspace{14mm}
\epsilon_{r}
\epsilon_{q}
Q_a(\epsilon_{r})
Q^*_a(\epsilon_{p})
Q_b(\epsilon_{q})
Q^*_b(\epsilon_{p})
+
\nonumber \\
& \hspace{16mm}
\epsilon_{q}
\epsilon_{p}
Q_a(\epsilon_{q})
Q^*_a(\epsilon_{r})
\left|Q_b(\epsilon_{p})\right|^2
+
\nonumber \\
& \hspace{18mm}
\epsilon_{r}
\epsilon_{p}
Q_a(\epsilon_{r})
Q^*_a(\epsilon_{q})
\left|Q_b(\epsilon_{p})\right|^2
\Bigr]
\Biggr\}
\nonumber
\end{align}

\noindent and for $D)$

\begin{align}
\label{eq:D}
\delta
=
\frac{1}{2}N^2\sum_{\substack{p,q=0\\p\neq q}}^{N-1}
\Biggl\{&
\frac{-1}{(\epsilon_p-\epsilon_q)^2}\times
\\
&
\Bigl[
\epsilon_{p}
\epsilon_{q}
\left|Q_a(\epsilon_p)
Q_b(\epsilon_q)\right|^2
+
\nonumber \\
& \hspace{7mm}
\epsilon_{p}^2
Q_a(\epsilon_p)
Q^*_a(\epsilon_q)
Q_b(\epsilon_p)
Q^*_b(\epsilon_q)
+
\nonumber \\
& \hspace{14mm}
\epsilon_{p}
\epsilon_{q}
Q_a(\epsilon_p)
Q^*_a(\epsilon_q)
Q_b(\epsilon_q)
Q^*_b(\epsilon_p)
\Bigr]\Biggr\}
\nonumber	
\end{align}

Summarizing, we can write the sum of squares corresponding to cross-correlations terms of the $\ISL$ as 
\begin{equation*}
\XX_{ab}=\frac{1}{2N} \sum_{j=0}^{N-1}\left|Q_a(\epsilon_j)Q^*_b(\epsilon_j)\right|^2 + \frac{16}{N^4}\left(\alpha+\beta+\gamma+\delta\right)
\end{equation*}

\noindent where the quantities $\alpha$, $\beta$, $\gamma$, $\delta$ are defined in \eqref{eq:A}, \eqref{eq:B}, \eqref{eq:C}, \eqref{eq:D}.

With the method presented above in conjunction with the method presented in \cite{Hoholdt_1988}, we can have an analytical expression for the $\ISL$ for any set of real sequences of odd length. The computation of the above equations seems to be hard at a first look, but in a number of cases, in particular for sequences from difference sets \cite{Hoholdt_2006} may lead to significant results.

In the following, we use this method to evaluate the asymptotic trend of the $\ISL$ of a set of sequences made up by different Rotations of a Legendre Sequence (RLS set) when $N$ grows to infinity.

\section{Legendre Sequences}
\label{sec:LegSec}

The \textit{Legendre Sequence} (LS) $\ell_0, \dots ,\ell_{N-1}$ exists for any prime $N$ and is defined as

\begin{align*}
\label{eq:LS}
	& \ell_0 = 1 \notag\\
	& \ell_j = 
	\begin{cases}
		1  & \text{if $j$ is a square $\pmod N$}\\ 
		-1 & \text{if $j$ is a nonsquare  $\pmod N$}
	\end{cases}
\end{align*}

A LS may be cyclically rotated $t_a$ positions to the left to obtain a \textit{Rotated Legendre Sequence} (RLS) $a_j$ defined as
\begin{equation*}
\label{eq:RRR}
	a_j= \ell_{j+t_a \pmod N} = \ell_{j+ f_a N  \pmod N}
\end{equation*}

\noindent with $f_a = t_a/N \in[0,1]$.

The asymptotic value of $\XX_{aa}$ for the family of RLS was calculated in
\cite{Golay_1983} and \cite{Hoholdt_1988}
\footnote{The first contribution relies on a ``Postulate of Mathematical Ergodicity'' to arrive at a result which is formally proved by the second.} 
noting that the asymptotic value of the modulus of the generating function of the LS ($|Q_\ell(\epsilon_j)|$) is independent of $j$,
yielding

\begin{equation}
\label{eq:AXXaa}
	\frac{\XX_{aa}}{N^2}
	\sim
	 \frac{2}{3} - 4\left|f_a - \frac{1}{2}\right| + 8\left(f_a - \frac{1}{2}\right)^2
\end{equation}

We follow the same path as in \cite{Hoholdt_1988} but for the calculation of the cross-correlations terms of the $\ISL$ $\XX_{ab}$.

To proceed, remember that the generating function of the LS is

\begin{equation}
\label{eq:gfLS}
Q_\ell(\epsilon_j)=\begin{cases}
1+\ell_j\sqrt{N} & \text{if $j\neq 0$ and $N=1\pmod 4$}\\
1+\iu\ell_j\sqrt{N} & \text{if $j\neq 0$ and $N=3 \pmod 3$}\\
1 & \text{if $j=0$}
\end{cases}
\end{equation}

Moreover, if we denote by $Q_a(\epsilon_j)$ the generating function of the RLS $a_j=\ell_{j+t_a \pmod N}$, then

\begin{equation*}
Q_a(\epsilon_j)=
\epsilon_j^{-t_a}Q_\ell(\epsilon_j)
\end{equation*}

Assume now that the two sequences $a_j$ and $b_j$ are obtained by rotating $\ell_j$ by, respectively, $t_a$ and $t_b$ positions to the left.
We may compute $S'$ as

\begin{equation*}
S'=\sum_{j=0}^{N-1}\left|\epsilon_j^{-t_a}Q_\ell(\epsilon_j)\epsilon_j^{t_b}Q^*_\ell(\epsilon_j)\right|^2=
\sum_{j=0}^{N-1}\left|Q_\ell(\epsilon_j)\right|^4
\end{equation*}

\noindent from \eqref{eq:gfLS} we know immediately that $\left|Q_\ell(\epsilon_j)\right|^4 \sim N^2$, then $S'\aseq N^3$.
Let us now compute the asymptotic values of $\alpha$, $\beta$, $\gamma$ and $\delta$ in \eqref{eq:A}, \eqref{eq:B}, \eqref{eq:C}, \eqref{eq:D} for any pair of RLS.

\begin{list}{\labelitemi}{\leftmargin=1em}
\item
For $\alpha$ in \eqref{eq:A} we have
\end{list}

\begin{align*}
\alpha
&=
\frac{1}{16}\left(\frac{1}{3}N^4+\frac{2}{3}N^2\right)S'\\
&\aseq \frac{1}{48}N^7
\end{align*}

%
\begin{list}{\labelitemi}{\leftmargin=1em}
\item
For $\beta$ in \eqref{eq:B}
\end{list}

\begin{align*}
\beta
=
\frac{1}{8}N^2\sum_{\substack{p,q=0\\p\neq q}}^{N-1}
\Biggl\{&
\left(
\frac{\epsilon_q+\epsilon_p}{\epsilon_p(\epsilon_q-\epsilon_p)^2}
\right)
\times
\\
&
\Bigl[
\epsilon_p^2\left|Q_\ell(\epsilon_p)\right|^2\epsilon_{p-q}^{t_b}Q_\ell(\epsilon_p)Q^*_\ell(\epsilon_q)+
\\
& \hspace{7mm}
\epsilon_{p}\epsilon_{q}\left|Q_\ell(\epsilon_p)\right|^2\epsilon_{q-p}^{t_b}Q_\ell(\epsilon_q)Q^*_\ell(\epsilon_p)+
\\
& \hspace{10.5mm}
\epsilon_{p}^2\epsilon_{p-q}^{t_a}Q_\ell(\epsilon_p)Q^*_\ell(\epsilon_q)\left|Q_\ell(\epsilon_p)\right|^2+
\\
& \hspace{14mm}
\epsilon_{q}\epsilon_{p}\epsilon_{q-p}^{t_a}Q_\ell(\epsilon_q)Q^*_\ell(\epsilon_p)\left|Q_\ell(\epsilon_p)\right|^2
\Bigr]\Biggr\}\\
\sim
\frac{1}{8}N^2\sum_{\substack{p,q=0\\p\neq q}}^{N-1}
\Biggl\{&
\left(
\frac{\epsilon_q+\epsilon_p}{\epsilon_p(\epsilon_q-\epsilon_p)^2}
\right)
\times
\\
&
\Bigl(
N^2 \ell_p \ell_q \epsilon_p^2 \epsilon_{p-q}^{t_b}+
N^2 \ell_p \ell_q \epsilon_p \epsilon_q \epsilon_{q-p}^{t_b}+
\\
& \hspace{13mm}
N^2 \ell_p \ell_q \epsilon_p^2 \epsilon_{p-q}^{t_a}+
N^2 \ell_p \ell_q \epsilon_p \epsilon_q \epsilon_{q-p}^{t_a}
\Bigr)
\Biggr\}\\
=
\frac{1}{8}N^4\sum_{\substack{p,q=0\\p\neq q}}^{N-1}
\Biggl\{&
\left(
\frac{\ell_p \ell_q}{(1-\epsilon_{p-q})^2}
\right)
\times\\
&
\Bigl(
\epsilon_{p-q}^{t_b+1}+
\epsilon_{p-q}^{t_b+2}+
\epsilon_{p-q}^{1-t_b}+
\epsilon_{p-q}^{-t_b}+
\\
& \hspace{20mm}
\epsilon_{p-q}^{t_a+2}+
\epsilon_{p-q}^{t_a+1}+
\epsilon_{p-q}^{1-t_a}+
\epsilon_{p-q}^{-t_a}
\Bigr)\Biggr\}\\
\end{align*}

\begin{multline*}
=
\frac{1}{8}N^4\sum_{\substack{k=-N+1\\k\neq 0}}^{N-1}
\left(
X_{\ell \ell}(k) + X_{\ell \ell}(N-k)
\right)\\
\frac{\epsilon_k^{t_b+1}+
\epsilon_k^{t_b+2}+
\epsilon_k^{1-t_b}+
\epsilon_k^{-t_b}+
\epsilon_k^{t_a+2}+
\epsilon_k^{t_a+1}+
\epsilon_k^{1-t_a}+
\epsilon_k^{-t_a}}
{(1-\epsilon_k)^2}
\end{multline*}

Note that $X_{\ell \ell}(k) + X_{\ell \ell}(N-k)$ is the periodic correlation \cite{Hoholdt_2006} of the LS. Then, from \cite{Golay_1983} and \cite{Golomb_2005} we know that $\left|X_{\ell \ell}(k) + X_{\ell \ell}(N-k)\right| \leq 3$ for Legendre sequences. Then, using the fact that $\sum_{k=1}^{N-1} \frac{1}{|1-\epsilon_k|^2} = O(N^2)$ (see \eqref{eq:D_smallAngle} and \eqref{eq:ReLi2_t} below and set $t=0$), and using the triangle inequality we get that $\beta = O(N^6)$.

%

\begin{list}{\labelitemi}{\leftmargin=1em}
\item
For the calculation of $\gamma$ in \eqref{eq:C}, following the same steps we did for $\beta$ we have
\end{list}

\begin{align*}
\gamma
\sim
\frac{1}{4}N^4
\sum_{\substack{p,q,r=0\\p\neq q\neq r}}^{N-1}
\Biggl\{
&
-\frac{\ell_q \ell_r}{(1-\epsilon_{p-q})(1-\epsilon_{p-r})}
\Bigl(
\epsilon_{p-r} \epsilon_{q-r}^{t_b}+
\\
&
\epsilon_{p-q} \epsilon_{q-r}^{-t_b}+
\epsilon_{p-r}^{t_b+1} \epsilon_{p-q}^{t_a+1}+
\epsilon_{p-r}^{t_a+1} \epsilon_{p-q}^{t_b+1}+
\\
& \hspace{6mm}
\epsilon_{p-r}^{-t_b} \epsilon_{p-q}^{t_a+1}+
\epsilon_{p-r}^{t_a+1} \epsilon_{p-q}^{-t_b}+
\epsilon_{p-q}^{-t_a} \epsilon_{p-r}^{-t_b}+
\\
& \hspace{12mm}
\epsilon_{p-r}^{-t_a} \epsilon_{p-q}^{-t_b}+
\epsilon_{p-r} \epsilon_{q-r}^{t_a}+
\epsilon_{p-q} \epsilon_{q-r}^{-t_a}
\Bigr)
\Biggr\}\\
=
\frac{1}{4}N^4
\sum_{\substack{u,v=-N+1\\u\neq v\neq 0}}^{N-1}
\Biggl\{&
-\frac{
X_{\ell \ell}(v-u) + X_{\ell \ell}(N-(v-u))
}{(1-\epsilon_{v})(1-\epsilon_{u})}
\Bigl(
\epsilon_{u} \epsilon_{u-v}^{t_b}+
\\
&
\epsilon_{v} \epsilon_{u-v}^{-t_b}+
\epsilon_{u}^{t_b+1} \epsilon_{v}^{t_a+1}+
\epsilon_{u}^{t_a+1} \epsilon_{v}^{t_b+1}+
\\
& \hspace{10mm}
\epsilon_{u}^{-t_b} \epsilon_{v}^{t_a+1}+
\epsilon_{u}^{t_a+1} \epsilon_{v}^{-t_b}+
\epsilon_{v}^{-t_a} \epsilon_{u}^{-t_b}+
\\
& \hspace{20mm}
\epsilon_{u}^{-t_a} \epsilon_{v}^{-t_b}+
\epsilon_{u} \epsilon_{u-v}^{t_a}+
\epsilon_{v} \epsilon_{u-v}^{-t_a}
\Bigr)
\Biggr\}
\\
\end{align*}

and again we have that $\gamma = O(N^6)$

%

\begin{list}{\labelitemi}{\leftmargin=1em}
\item
For $\delta$ in \eqref{eq:D} we have
\end{list}

\begin{align*}
\delta
=&
\frac{1}{2}N^2\sum_{\substack{p,q=0\\p\neq q}}^{N-1}
\Biggl\{
\frac{-1}{(\epsilon_p-\epsilon_q)^2}\times
\Bigl[
\epsilon_{p}
\epsilon_{q}
\left|Q_\ell(\epsilon_p)
Q_\ell(\epsilon_q)\right|^2
+
\\
& \hspace{18mm}
\epsilon_{p}^2
\epsilon_{p-q}^{t_a}
Q_\ell(\epsilon_p)
Q^*_\ell(\epsilon_q)
\epsilon_{p-q}^{t_b}
Q_\ell(\epsilon_p)
Q^*_\ell(\epsilon_q)
+
\\
& \hspace{18mm}
\epsilon_{p}
\epsilon_{q}
\epsilon_{p-q}^{t_a}
Q_\ell(\epsilon_p)
Q^*_\ell(\epsilon_q)
\epsilon_{q-p}^{t_b}
Q_\ell(\epsilon_q)
Q^*_\ell(\epsilon_p)
\Bigr]\Biggr\}\\
\sim&
-
\frac{1}{2}N^4\sum_{\substack{p,q=0\\p\neq q}}^{N-1}
\Biggl\{
\frac{\epsilon_{q-p} +
\epsilon_{q-p}^{-t_a-t_b}+
\epsilon_{q-p}^{1-t_a+t_b}}
{(1-\epsilon_{q-p})^2}
\Biggr\}\\
=&
-
\frac{1}{2}N^4\sum_{\substack{k=-N+1\\k\neq 0}}^{N-1}
\frac{
\Bigl(
\epsilon_{k} +
\epsilon_{k}^{-t_a-t_b}+
\epsilon_{k}^{1-t_a+t_b}
\Bigr)
}{(1-\epsilon_{k})^2}
(N-|k|)
\\
=&
-
N^4\sum_{k=1}^{N-1}
\frac{
\Bigl(
\epsilon_{k} +
\epsilon_{k}^{-t_a-t_b}+
\epsilon_{k}^{1-t_a+t_b}
\Bigr)
}{(1-\epsilon_{k})^2}
(N-|k|)
\\
\end{align*}

Larger values of the summand are those for $k$ close to 1, which make the denominator close to zero and numerator $\sim cN$ for some constant $c$ (for $k$ close to $N-1$, the denominator becomes also close to zero but the numerator is $O(1)$).

Exploiting this and using the small angle approximation for the complex exponential, we may write

\begin{align}
\label{eq:D_smallAngle}
\delta
\aseq &
-N^5
\sum_{k=1}^{N-1}\frac{\epsilon_k+\epsilon_k^{-t_a-t_b}+\epsilon_k^{1-t_a+t_b}}{-\frac{4\pi^2}{N^2}k^2}
\end{align}

To continue, we recall the definition of the Dilogarithm function and its series expansion valid for $\left|z\right| \leq 1$

\begin{equation}
\label{eq:Di2}
{\rm Li}_2
\left(z
\right)
=
-\int_0^1 \frac{ln(1-zt)}{t}dt
=
\sum_{k=1}^{\infty}\frac{z^k}{k^2}
\end{equation}

Taking the real part of \eqref{eq:Di2} and evaluating on the unit circle gives \cite[eq. (8.7)]{Maximon_2003}

\begin{equation}
\label{eq:ReLi2}
\Re{
{\rm Li}_2
\left(
e^{\iu\theta}
\right)
}
=
\Re{
\sum_{k=1}^{\infty}\frac{e^{\iu k \theta}}{k^2}
}
=
\frac{1}{6}\pi^2-
\frac{1}{4}\left|\theta\right| \left(2 \pi - \left|\theta\right| \right)
\end{equation}

Exploiting \eqref{eq:ReLi2} and concentrating on the first period $0\le\frac{t}{N}\le 1$ we obtain

\begin{equation}
\label{eq:ReLi2_t}
\Re{
\sum_{k=1}^{\infty}\frac{\epsilon_k^t}{k^2}
}
=
\pi^2\left[\frac{1}{6}-
\umod{\frac{t}{N}}
\left(
1-
\umod{\frac{t}{N}}
\right)
\right]
\end{equation}

\noindent where $\umod{\cdot}=\cdot \pmod 1$.

\newcommand{\modpar}[1]{\umod{#1}\left(1-\umod{#1}\right)}
\newcommand{\asLi}[1]{\frac{1}{6}-\modpar{#1}}

Hence, since we know that $\delta$ is real
\begin{align*}
\delta
\aseq &
\frac{1}{4}N^7 \Biggl\{
\frac{1}{6}+\asLi{-\frac{t_a+t_b}{N}}+
\\
&
\asLi{\frac{t_b-t_a}{N}}
\Biggr\}\\
=&
\frac{1}{4}
N^2\Bigl\{
\frac{1}{2}-\modpar{-f_a-f_b}-
\\
&
\modpar{f_b-f_a}
\Bigr\}\\
=&
\frac{1}{4}
N^2\Bigl\{
\frac{1}{2}-\modpar{f_a+f_b}-
\\
&
\modpar{f_a-f_b}
\Bigr\}\\
\end{align*}

\noindent where we have defined $f_a=\frac{t_a}{N}$ and $f_b=\frac{t_b}{N}$. Then, exploiting the symmetries of a quadratic form of a modulus function we have for $0\le f_a,f_b\le 1$
\begin{align*}
\modpar{f_a+f_b} & = \frac{1}{4}-\left(\left|f_a+f_b-1\right|-\frac{1}{2}\right)^2\\
\modpar{f_a-f_b} & = \frac{1}{4}-\left(\left|f_a-f_b\right|-\frac{1}{2}\right)^2
\end{align*}

\noindent so that
\[
\delta
\aseq \frac{1}{4}N^7\left[\left(\left|f_a+f_b-1\right|-\frac{1}{2}\right)^2+\left(\left|f_a-f_b\right|-\frac{1}{2}\right)^2\right]
\]

Based on the above we are now interested in computing the asymptotic value of

\begin{align}
\label{eq:AXXab}
\frac{1}{N^2}\XX_{ab}
= &
\frac{1}{2N^3}\left(S'+S''\right)
\sim
\frac{1}{2N^3}\left[N^3+\frac{16}{N^4}\left(\alpha+\beta+\gamma+\delta\right)\right]\notag
\\
\sim &
\frac{2}{3}
+
2 \left(\left|f_a+f_b-1\right|-\frac{1}{2}\right)^2 + 2 \left(\left|f_a-f_b\right|-\frac{1}{2}\right)^2
\end{align}

Going back to our original problem for calculation of the $\ISL$ value of a set of M sequences $x^{(p)}_j$ with $j = 0,\dots, N-1$ and $p = 0,\dots, M-1$, where each $x^{(p)}$ is made by a different rotation $f_p$ of a LS (RLS set), replacing \eqref{eq:AXXaa} and \eqref{eq:AXXab} into \eqref{eq:ISL_sim} we finally have that

\begin{multline}
\label{eq:ISLM}
\frac{\ISL}{N^2}
\sim 
	\sum_{p=0}^{M-1}
	\frac{2}{3} - 4\left|f_p - \frac{1}{2}\right| + 8\left(f_p - \frac{1}{2}\right)^2 +\\
  \sum_{p=0}^{M-1}\sum_{\substack{q=0\\p\neq q}}^{M-1} 
  \frac{2}{3} + 2 \left(\left|f_p+f_q-1\right|-\frac{1}{2}\right)^2 + \\
  2 \left(\left|f_p-f_q\right|-\frac{1}{2}\right)^2
\end{multline}

\begin{figure}
	\begin{center}
		\includegraphics[width=0.80\columnwidth]{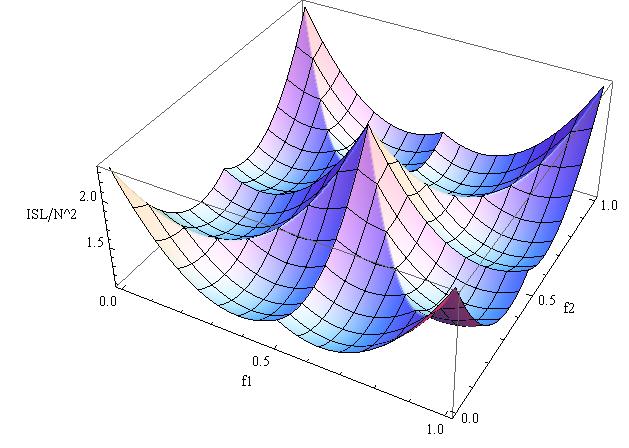} \\
		\hspace{10mm}(a) \\
		\includegraphics[width=0.64\columnwidth]{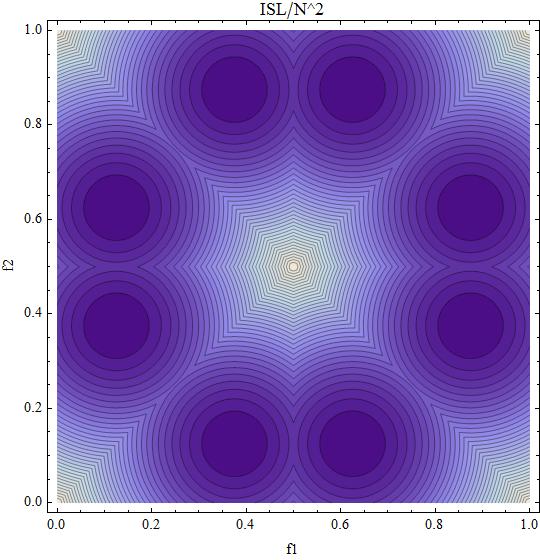} \\
		\hspace{5mm}(b) \\
	\end{center}
\caption{\label{fig:ISL2}Plots of $\ISL$ for $M=2$ as a function of $f_1$ and $f_2$: (a) 3D-view, (b) iso-$\ISL$ lines}
\end{figure}

\begin{figure*}
  \centering
  \begin{tabular}{ccc}
	  \includegraphics[width=0.64\columnwidth]{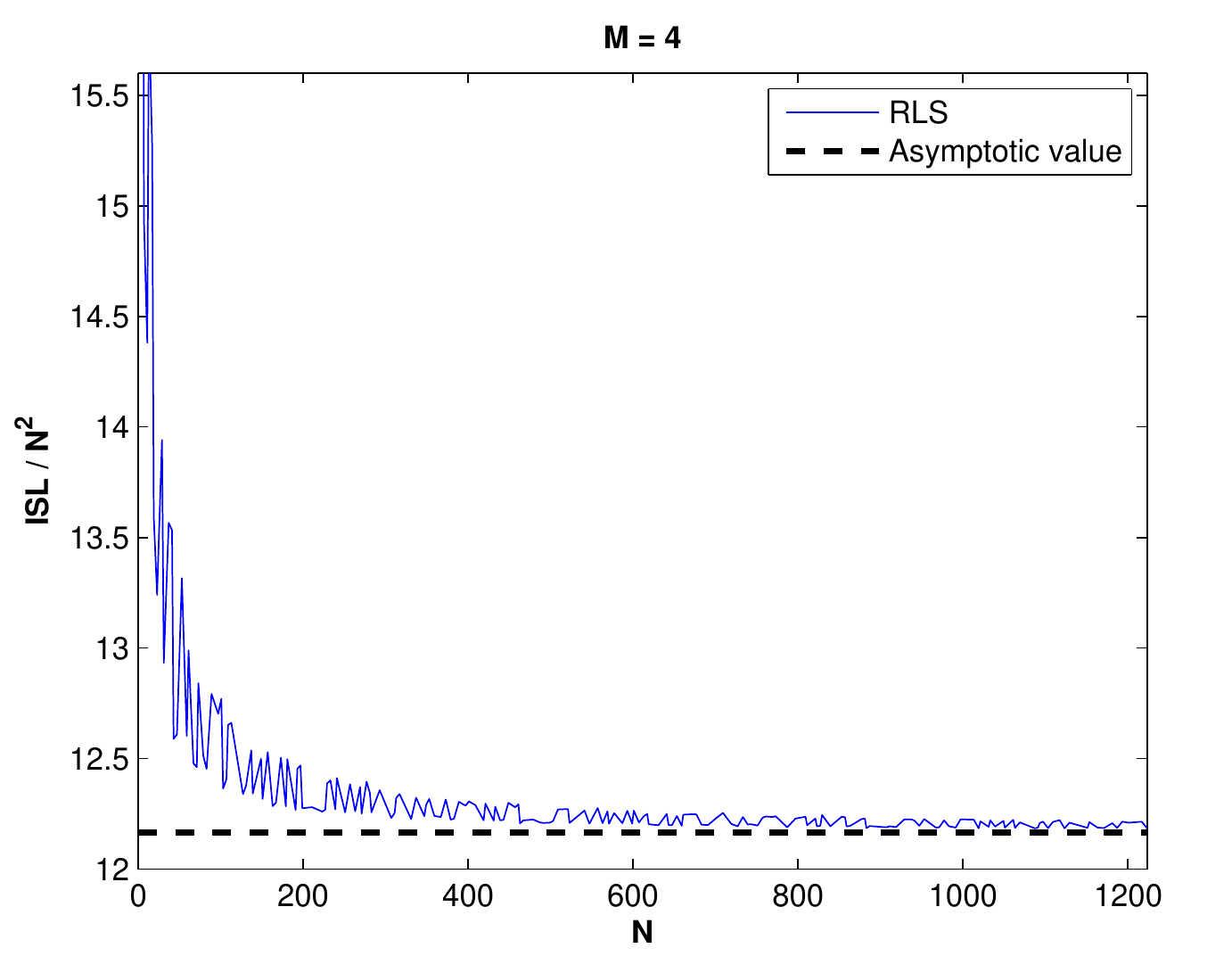} &
	  \includegraphics[width=0.64\columnwidth]{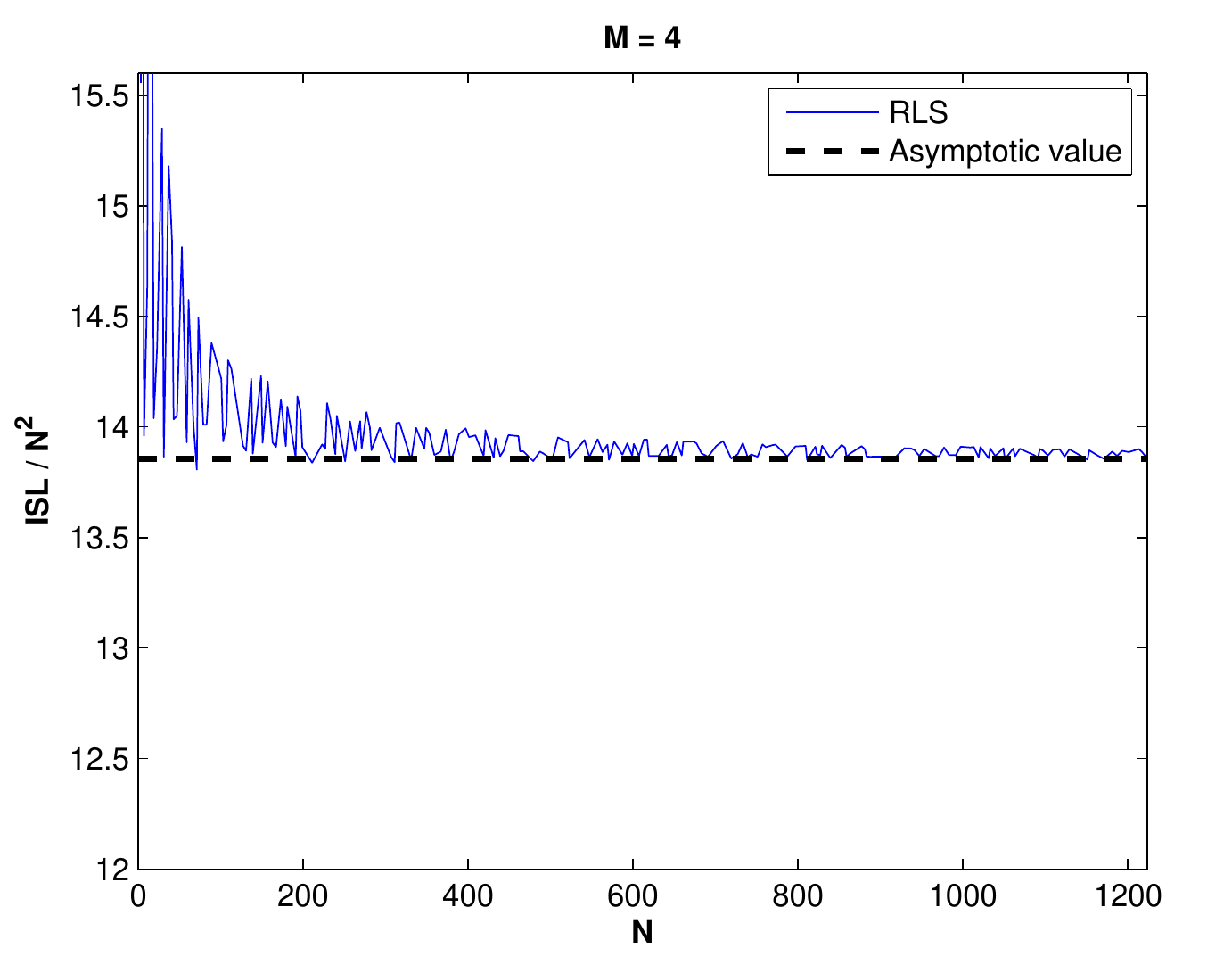} &
	  \includegraphics[width=0.64\columnwidth]{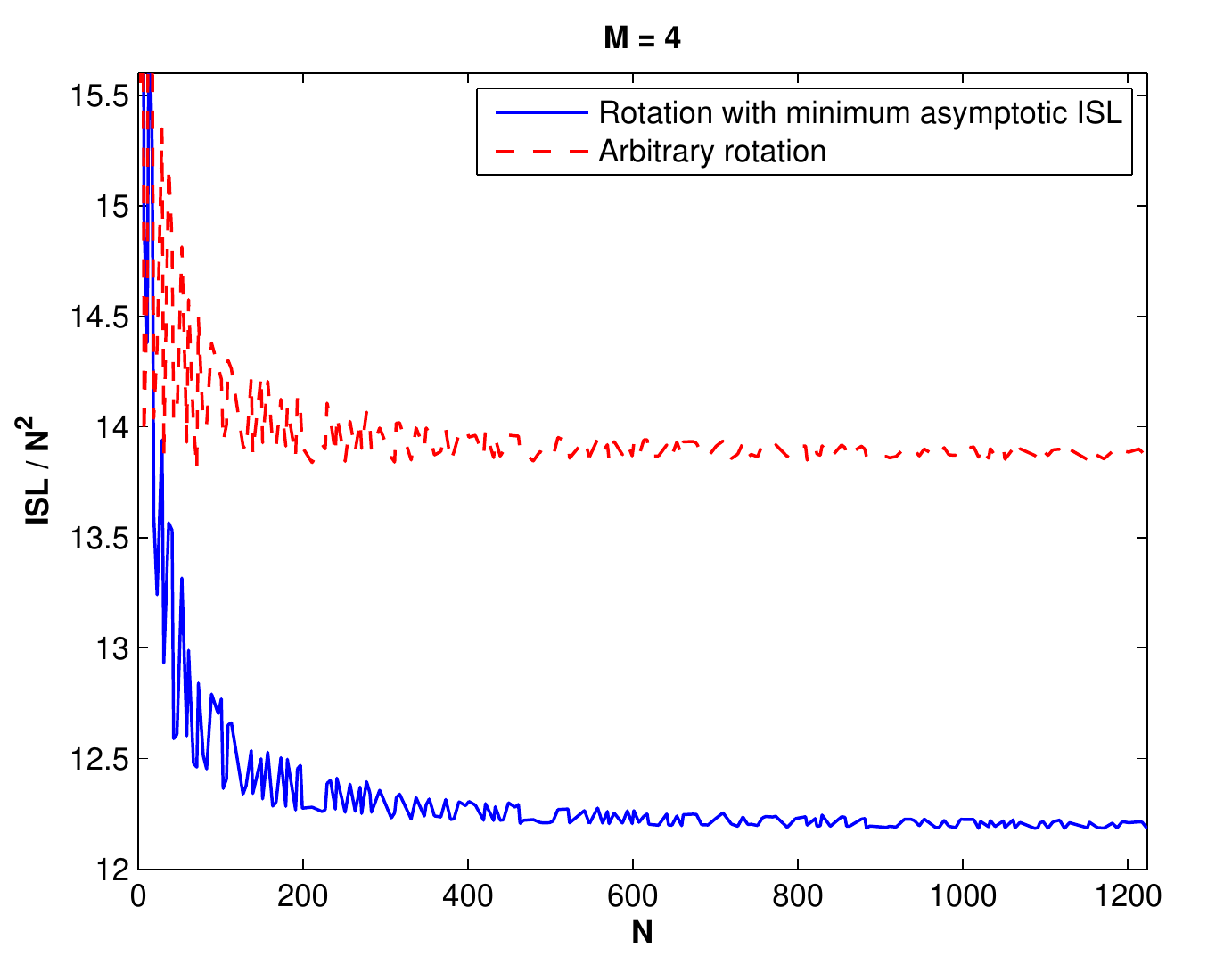} \\
	  (a) & (b) & (c)
	\end{tabular}
\caption{\label{fig:ISL_M4}Plots of $\ISL$ for $M=4$ as a function of N: (a) Rotations minimizing asymptotic $\ISL$, (b) Arbitrary rotations. In dotted line the asymptotic value. (c) Comparison of the cases in a) and b).}
\end{figure*}

As an example, Figure \ref{fig:ISL2} reports the 3D and contour plot
of the right-hand side of \eqref{eq:ISLM} for $M=2$.
Direct visual inspection of that Figure confirms that minima exists and can be easily identified.
We offered a preliminary exploit of the result in \cite{Haboba_2010}
where an optimization procedure was developed to find the optimal rotations that minimize the $\ISL$ for any sequences length $N$.

As another example, in Figure \ref{fig:ISL_M4} we plot the $\ISL$ for $M=4$ as a function of the sequence length $N$. In case a) the values of rotation are those that minimize the asymptotic $\ISL$, while in case b) we use an arbitrary rotation. In both cases we can see that the trend of the plots is in agreement with the asymptotic value calculated.
In the same figure in case c), we plot together both curves in a) and b) to show that the one that achieves the minimum asymptotic value of $\ISL$, also achieves the minimum $\ISL$ value for sequences length greater than approximately $20$. For different choices of rotations and different number of sequences ($M$), the behavior is the same than presented.

\section{Conclusion}
\label{sec:conc}

We apply a method based on generating functions, which has already been proposed for the calculation of the $\ISL$ of a sequence, to the calculation of the cross-correlation components of the $\ISL$ of a set of sequences.

The apparent complexity of the resulting expressions can be tackled in the asymptotic conditions for sequences whose generating function has a relatively simple trend.

Since this is the case of Legendre sequences, we are able to derive an analytical expression for the asymptotic $\ISL$ of sets of rotated Legendre sequences.

Such an expression can be exploited to drive the optimization procedure needed to construct small-$\ISL$ sets of antipodal sequences with potential applications to communication and active sensing systems.

\bibliographystyle{IEEEbib}
\bibliography{refs}

\end{document}